\documentclass[twocolumn,aps,prl,preprintnumbers,showpacs,nofootinbib]{revtex4}
\usepackage[utf8]{inputenc}
\usepackage[english]{babel}
\usepackage{amsmath}
\usepackage{amsfonts}
\usepackage{amssymb}
\usepackage{epsfig}
\usepackage{graphics,psfrag,rotating}
\usepackage{graphicx}
\usepackage{dcolumn}
\usepackage{bm}
\usepackage{mathtools,amssymb,lipsum}
\usepackage{epstopdf}
\usepackage{color}
\usepackage[usenames,dvipsnames,svgnames]{xcolor}
\usepackage[colorlinks=true,
            linkcolor=red,
            urlcolor=gray,
            citecolor=blue]{hyperref}
  \usepackage{hyperref}

\newcommand{\lsim}   {\mathrel{\mathop{\kern 0pt \rlap
{\raise.2ex\hbox{$<$}}}
 \lower.9ex\hbox{\kern-.190em $\sim$}}}
\newcommand{\gsim}   {\mathrel{\mathop{\kern 0pt \rlap
{\raise.2ex\hbox{$>$}}}
\lower.9ex\hbox{\kern-.190em $\sim$}}}

\def\3nab{\tilde{\nabla}}

\def\hsp5{\hspace{5mm}}

\def\case#1/#2{\textstyle\frac{#1}{#2}}

\def\ber {\begin{eqnarray}}
\def\eer {\end{eqnarray}}
\def\bea {\begin{eqnarray}}
\def\eea {\end{eqnarray}}

\def\bc {\begin{center}}
\def\ec {\end{center}}
\def\case#1/#2{\frac{#1}{#2}}

\newcommand{\bw}{\begin{widetext}}
\newcommand{\ew}{\end{widetext}}

\newcommand{\be}{\begin{equation}}
\newcommand{\bse}{\begin{subequation}}
\newcommand{\ese}{\end{subequation}}
\newcommand{\ee}{\end{equation}}
\newcommand{\eei}{\end{eqnarray}\indent\indent}
\newcommand{\ba}{\begin{array}}
\newcommand{\ea}{\end{array}}
\newcommand{\bal}{\begin{eqnarray}}
\newcommand{\eal}{\end{eqnarray}}

\def\case#1/#2{\textstyle\frac{#1}{#2} }

\newcommand{\bver}{\begin{verbatim}}
\newcommand{\ever}{\end{verbatim}}

\begin{document}

%\preprint{APS/123-QED}

\title{ Test of Barrow entropy using a model independent approach}
%{Density perturbations in the Jordan-Fierz-Brans-Dicke theory}%
\author{
A. Salehi\footnote{Email: salehi.a@lu.ac.ir},
}
\affiliation{ Department of Physics, Lorestan University, Khoramabad, Iran}

\date{\today}% It is always \today, today,

\begin{abstract}
Taking into consideration of a fractal structure for the black hole
horizon, Barrow argued that the area law of entropy get
modified due to quantum-gravitational effects. Accordingly, the corrected entropy
takes the form $S\sim A^{1+\frac{\Delta}{2}}$, where $0\leq\Delta\leq1,$ indicates
the amount of the quantum-gravitational deformation effects. By considering the modified Barrow entropy associated with the apparent horizon, the Friedmann equations get modified as well. We show that considering a universe filled with the matter and cosmological constant $\Lambda$, it is possible to determine the amount of deviation from standard cosmology by reconstructing the parameter $\delta$ in terms of curvature parameters $\{q,Q,\Omega_{k}\}$ as $\Delta=\frac{(Q-1-\Omega_k)(1+\Omega_k)}{(1+\Omega_k+q)^{2}}$.
Here, $q$ is the deceleration parameter and $Q$ is the third derivative of scale factor . This relation provides some advantages. The first is that it indicates that there is profound connection between quantum-gravitational deformation effects and curvature effects, for $\Omega_k\simeq0$ the pair $\{q,Q\}$ can be regarded as deviation curvature factors which reflect the amount of deviation of the model from the standard model. The second interesting feature is that, since this pair are observational parameters which can be directly measured in a model independent approach, they can be regarded as powerful tools to enable us to put constraint on parameter $\Delta$ and test the Barrow entropy model. Our analysis predicts the value for $Q_{0}$ which is slightly deviates from 1 as $(Q_{0}-1)<0.001$. This can be a relativity well target and criterion for theoretical and observational measurements of parameter $Q_{0}$. Hence we can hope and wait the improvement of the high redshift data in the future to support it.
\end{abstract}


%04.50.Kd	Modified theories of gravity
%04.25.Nx 	Post-Newtonian approximation; perturbation theory; related approximations
% 95.36.+x 	Dark energy (see also 98.80.-k Cosmology)
% 98.80.-k	 Cosmology (see also section 04 General relativity and gravitation; for origin and evolution of galaxies, see 98.62.Ai; for elementary particle and nuclear processes, see 95.30.Cq; for dark matter, see 95.35.+d; for dark energy, see 95.36.+x; for superclusters and large-scale structure of the Universe, see 98.65.Dx)
%
%
% 04.50.-h 	Higher-dimensional gravity and other theories of gravity (see also 11.25.Mj Compactification and four-dimensional models, 11.25.Uv D branes)
%04.20.-q	         Classical general relativity (see also 02.40.-k Geometry, differential geometry, and topology)
%04.20.Cv	Fundamental problems and general formalism
%04.50.-h	         Higher-dimensional gravity and other theories of gravity (see also 11.25.Mj Compactification and four-dimensional models, 11.25.Uv D branes)
% 98.80.Jk        Mathematical and relativistic aspects of cosmology
% 04.20.Jb        Exact solutions


\maketitle
The quantum phenomenon of Hawking radiation indicates that black hole has a temperature proportional to its surface gravity and an entropy proportional to its horizon area \cite{Shey}-\cite{Bekenstein}. This issue led to discovery of profound connection between gravity and thermodynamics which was first addressed by Jacobson \cite{Xin} who disclosed that the Einstein gravitational theory for the spacetime metric can be extracted from the horizon entropy-area relation by using the fundamental Clausius relation  $\delta Q = T\Delta S$ \cite{Horava}. The investigations on the relation between Einstein field equations and first law of thermodynamics in the setup of black hole spacetime, have been generalized to the cosmological context to derive Friedmann equations with any spatial curvature by applying the Clausius relation to apparent horizon of the FRW universe \cite{Fischler}-\cite{cai2}. See \cite{cc} for further studies of thermodynamical aspects of gravity.
Recently, Barrow \cite{5} explained that quantum gravitational effects might create complex and fractal properties on the black hole horizon. In this perspective, the entropy of a black hole no longer follows the law of area, but can
instead be represented by exponential raising of the area $ \Delta $ as
 \begin{equation}\label{f1}
 S_B=\left( \dfrac{A}{A_0}\right)^{1+\frac{\Delta}{2}}
 \end{equation}
 Where $ A $ is the normal horizon region and $ A_0 $ is the Planck region. Quantum gravitational disturbance is shown by the new represented symbol $ \Delta $. There are characteristic values for $ \Delta $. For example, when $ \Delta=0 $, we have the simplest horizon. On the other hand, when $ \Delta=1 $, we see the phenomenon called maximum deformation.
 The authors of
 In this paper we want to show that there is connection between parameter $\Delta$ which reflects the quantum gravitational effects and deceleration parameter and jerk parameter.
  which arise from curvature and its changes.
Several studies have been devoted to the possibility that quantum gravity might triggered by curvature, but the relevant
literature has so far focused exclusively on a subclass of scenarios such that the quantum-gravity effects
are independent of (macroscopic) curvature.
 In this paper we want to show that there is connection between parameter $\Delta$ which reflects the quantum gravitational effects and curvature effects.
  which arise from curvature and its changes.
FLRW metric In the background of FRW universe, the line elements of the metric
\begin{equation}
ds^2=-dt^2+a^2(t)\left(\frac{dr^2}{1-kr^2}+r^2(d\theta^2+\sin^2\theta
d\phi^2)\right),
\end{equation}
where $a(t)$ is scale factor of the universe, $ k = 1, 0, -1$ stand for closed, flat and open geometries
respectively,
 and $(t,r,\theta, \phi)$ are the
co-moving coordinates.In this we have adopted the convention of
$a_{0} = 1$, where the subscript 0 denotes the value at present
time (zero redshift).
 The Ricci scalar curvature is given by
 \begin{align}
 R=-6\Big(\dot{H}+2H^{2}+\frac{k}{a^{2}}\Big)
 \end{align}
 where, $H =\frac{\dot{a}}{a}$ is the Hubble parameter, dot denotes the
derivative with respect to the cosmic time $t$. The deceleration parameter $q$ and the third derivative of dimensionless scale factor $Q$
 defined as
 \begin{align}
 &q=-\frac{\ddot{a}}{aH^{2}}=-\Big(1+\frac{\dot{H}}{H^{2}}\Big)\\
 &Q=\frac{\dddot{a}}{aH^{3}}=\frac{\ddot{H}}{H^{3}}-3q-2
 \end{align}
 One can obtain $q$ and $Q$ in terms of $R$ and its time derivation $\dot{R}$ as
   \begin{align}\label{q12}
& q=\frac{R}{6H^{2}}+1+\Omega_{k}\\
& Q=-\frac{\dot{R}}{6H^{3}}-\frac{R}{6H^{2}}+3\Omega_{k}+3\label{q13}
  \end{align}
Where, $\Omega_{k}=\frac{k}{a^{2}H^{2}}$. The equations (\ref{q12}) and (\ref{q13}) indicate that there is connection between parameters $(q,Q)$ and curvature $R$ and its time derivation $\dot{R}$. Hence in the rest of the paper we may call the as curvature parameters.
In principal by considering the modified Barrow entropy associated with the apparent horizon, the Friedmann equations get modified as well. In this paper we show that it is possible to determine the amount of deviation from standard cosmology by reconstructing the parameter $\Delta$ in terms of curvature parameters $\{q,Q\}$.

\section{Modified Friedmann Equations based on Barrow entropy}\label{FIRST}
In this section we present derivation of modified Friedmann equations based on Barrow entropy by using
the gravity-thermodynamics conjecture. Modification of Friedman equations based on Barrow entropy was first explored by \cite{Emmanuel}. Then another approach was presented by \cite{SheBFE}. Although these two approaches seem different, we show that they are equivalent. The
author of \cite{Emmanuel} deduced that in an expanding universe, during a time interval $dt$ the heat flow through the horizon
is easily found to be \cite{ca}
\begin{equation}\label{de0}
\delta Q=-dE=A(\rho_m+p_m)H r_{A}dt.
\end{equation}
Where $A=4\pi r_{A}^{2}$ and $r_{A}^{2}$ is radius of apparent horizons. From the thermodynamical viewpoint the
apparent horizon is a suitable horizon consistent with first and
second law of thermodynamics \cite{wang1}-\cite{sheyECFE}. Assuming
that the Universe is bounded by the apparent horizon
of radius\begin{equation}\label{FRWapphor22}
\tilde{r}_A={1}/{\sqrt{H^2+k/a^2}}
\end{equation}\label{FRWapphor}
 and considering that for the universe horizon the temperature associated with the horizon is given by
\cite{Padmanabhan}
\begin{equation}\label{T}
T_h=\frac{1}{2 \pi \tilde r_A}
\end{equation}
We describe the content of the Universe as a perfect
fluid of energy-momentum tensor $
T_{\mu\nu}=(\rho_{m}+p_{m})u_{\mu}u_{\nu}+p_{m}g_{\mu\nu}$, where $\rho_{m}$ and
$p_{m}$ are the energy density and pressure, respectively. The
energy-momentum tensor is conserved, $\nabla_{\mu}T^{\mu\nu}=0$,
which implies the continuity equation, $\dot{\rho_{m}}+3H(\rho_{m}+p_{m})=0$.
Differentiating the Barrow entropy(\ref{f1}), yields
\begin{eqnarray} \label{dS}
dS_h=(2+\Delta)\left(\frac{4\pi}{A_{0}}\right)^{1+\Delta/2}
 {\tilde
{r}_{A}}^{1+\Delta} \dot{\tilde {r}}_{A} dt.
\end{eqnarray}
  Inserting these relations into the first law of
thermodynamics, and substituting  $\dot{{r}}_A$ using
(\ref{FRWapphor22}), we finally result to
\begin{align}\label{FRWgfe1}
&-(4\pi)^{(1-\Delta/2)}A_{0}^{(1+\Delta/2)}(\rho_m+p_m)=\nonumber\\ &2(2+\Delta)
\frac{\dot{H}-\frac{k}{a^2}}{\left(H^2+\frac{k}{
a^2}\right)^{\Delta/2}}.
\end{align}
Lastly,
integrating, for the validity region
$0\leq\Delta\leq1$ gives
 \begin{align}\label{firs}
\frac{ (4\pi)^{(1-\Delta/2)}A_{0}^{(1+\Delta/2)} }{6} \rho_m=&\frac{2+\Delta
}{2-\Delta} \left(H^2+\frac{k}{a^2}\right)
^{1-\Delta/2}\nonumber &\\-\frac{{C}}{3} A_{0}^{(1+\Delta/2)},
\end{align}
With C the integration constant.\
In the second approach which explored by \cite{SheBFE}, the temperature associated with the horizon is given by
\cite{Cai1}
\begin{equation}\label{T}
T_h=-\frac{1}{2 \pi \tilde r_A}\left(1-\frac{\dot {\tilde
r}_A}{2H\tilde r_A}\right),
\end{equation}

Since the Universe is expanding, the work density
associated with the volume change of the expanding universe, is
also given by $W=(\rho-p)/2$ \cite{Hay2}. Using the gravity-thermodynamics conjecture, the Friedmann equation can be obtained by
considering the Universe as a thermodynamic system
bounded by the apparent horizon and applying the first
law of thermodynamics
\begin{equation}\label{FL}
dE = T_h dS_h + WdV,
\end{equation}
Note that the $dE$ which defied here is different from that defined in the equation (\ref{de0}) of the first approach.  In equation (\ref{de0}) the $dE$  is just the energy flux crossing
the apparent horizon, and the apparent horizon radius
is kept fixed during an infinitesimal internal of time $dt$. Also, since, it was assumed that for infinitesimal internal $\dot{r_{A}}=0$, then the definition of temperature in two approach are different, in other word in the first approach, the term related to $\dot{r_{A}}=0$ has bee omitted .
However, in the equation (\ref{FL}) of the second, it was assumed that the first law
of thermodynamics on the apparent horizon in the form,
$dE$ is the change in the
energy inside the apparent horizon due to the volume
change $dV$ of the expanding Universe
where $E=\rho V$ is the total energy of the Universe of 3-dimensional volume $V=\frac{4\pi}{3}\tilde{r}_{A}^{3}$ with the area of apparent horizon
$A=4\pi\tilde{r}_{A}^{2}$ and $T_{h}$ and $S_{h}$ are, respectively,
the temperature and entropy associated with the apparent horizon.
 Taking differential form of the total
matter and energy, we find
\begin{equation}
dE=4\pi\tilde
 {r}_{A}^{2}\rho d\tilde {r}_{A}+\frac{4\pi}{3}\tilde{r}_{A}^{3}\dot{\rho} dt
 \end{equation}
By combining with conservation equation, we obtain
\begin{equation}
\label{dE2}
 dE=4\pi\tilde
 {r}_{A}^{2}\rho d\tilde {r}_{A}-4\pi H \tilde{r}_{A}^{3}(\rho_{m}+p_{m}) dt.
\end{equation}
%%%%%%%%%%%%%%%%%%%%%%%%%%%%%%%%%%%%%%%%%%%%%%%%%%%%%%%%%%%%%%%%%%%%%%%%%%%%%%%%%%%%%%

Finally, combining Eqs. (\ref{T}), (\ref{dE2}) and (\ref{dS}) with
the first law of thermodynamics (\ref{FL}) and using continuity relation, after some algebraic
calculations, we obtain
\begin{equation} \label{Fried2}
-\frac{2+\Delta}{2\pi A_0
}\left(\frac{4\pi}{A_0}\right)^{\Delta/2} \frac{d\tilde
{r}_{A}}{\tilde {r}_{A}^{3-\Delta}}=
 \frac{d\rho_{m}}{3}.
\end{equation}
After integration, we find the first modified Friedmann equation
in Barrow cosmology,
\begin{equation} \label{Fried4}
\left(H^2+\frac{k}{a^2}\right)^{1-\Delta/2} = \frac{8\pi G_{\rm
eff}}{3} \rho_{m}+\frac{\Lambda}{3},
\end{equation}
where $\Lambda$ is a constant of integration which can be
interpreted as the cosmological constant, and  $G_{\rm eff}$
stands for the effective Newtonian gravitational constant,$G_{\rm eff}\equiv \frac{A_0}{4} \left(
\frac{2-\Delta}{2+\Delta}\right)\left(\frac{A_0}{4\pi
}\right)^{\Delta/2}$.
 Eq.
(\ref{Fried4}), can be rewritten as
\begin{equation} \label{Fried5}
\left(H^2+\frac{k}{a^2}\right)^{1-\Delta/2} = \frac{8\pi G_{\rm
eff}}{3}(\rho_{m}+\rho_{\Lambda}).
\end{equation}
Where, $\rho_{\Lambda}={\Lambda}/(8\pi G_{\rm eff})$.
The second Friedmann equation, can be obtained by the
continuity equation with the first Friedmann equation
(\ref{Fried4}).
\cite{SheBFE}
\begin{eqnarray}
&&(2-\Delta)\frac{\ddot{a}}{a}
\left(H^2+\frac{k}{a^2}\right)^{-\Delta/2}+(1+\Delta)\left(H^2+\frac{k}{a^2}\right)^{1-\Delta/2}
\nonumber
\\
&&=-8\pi G_{\rm eff}(p_{m}+p_{\Lambda}),\label{2Fried3}
\end{eqnarray}
where $p_{\Lambda}=-{\Lambda}/(8\pi G_{\rm eff})$. In the limiting case where
$\Delta=0$ ($G_{\rm eff}\rightarrow G$), Eq. (\ref{2Fried3})
reduces to the second Friedmann equation in standard cosmology.\\
It is easy to show that equations (\ref{firs}) and (\ref{FRWgfe1}) of the first approach are equivalent to the equations (\ref{Fried5}) and (\ref{2Fried3}) of the second approach. Note tat equation (\ref{FRWgfe1})can be simplified as
\begin{align}\label{fs}
 &\left(H^2+\frac{k}{a^2}\right)
^{1-\Delta/2}=(\frac{2-\Delta
}{2+\Delta})\frac{ (4\pi)^{(1-\Delta/2)}A_{0}^{(1+\Delta/2)} }{6} \rho_m \\ \nonumber &+ \frac{{C}}{3}(\frac{2-\Delta
}{2+\Delta}) A_{0}^{(1+\Delta/2)},
\end{align}
Equation (\ref{fs}) can be simplified as
\begin{equation} \label{Fried7}
\left(H^2+\frac{k}{a^2}\right)^{1-\Delta/2} = \frac{8\pi G_{
eff}}{3}\rho_{m}+\Lambda.
\end{equation}
Where, \begin{align}
&G_{eff}=(\frac{2-\Delta
}{2+\Delta})\frac{ (4\pi)^{(1-\Delta/2)}A_{0}^{(1+\Delta/2)} }{6}=\\ &\frac{A_0}{4} \left(
\frac{2-\Delta}{2+\Delta}\right)\left(\frac{A_0}{4\pi
}\right)^{\Delta/2}\end{align}
  and \begin{align}\Lambda=\frac{{C}}{3}(\frac{2-\Delta
}{2+\Delta}) A_{0}^{(1+\Delta/2)}
\end{align}
We see that equation (\ref{Fried7}) which was obtained in the first approach is exactly equivalent to the equation \ref{Fried5} or \ref{Fried4} in the second approach. It is also easy to show that from equation (\ref{FRWgfe1}), we can obtain equation (\ref{2Fried3}).
Note that equation (\ref{FRWgfe1}) can be rewritten as
\begin{equation}\label{new0}
  (2-\Delta)(\dot{H}-\frac{k}{a^{2}})(\left(H^2+\frac{k}{a^2}\right)^{-\Delta/2}=-8\pi G_{\rm eff}(\rho_{m}+p_{m})
\end{equation}
Considering that $\dot{H}=\frac{\ddot{a}}{a}-H^{2}$. Hence, the equation (\ref{new0}) can be rewritten as
\begin{align}\label{new00}
  &(2-\Delta)(\frac{\ddot{a}}{a})\left(H^2+\frac{k}{a^2}\right)^{-\Delta/2} \nonumber\\&-(2-\Delta)\left(H^2+\frac{k}{a^2}\right)^{1-\Delta/2}=-8\pi G_{\rm eff}(\rho_{m}+p_{m})
\end{align}
Where, using equation (\ref{Fried5}) and noting that $p_{\Lambda}=-\rho_{\Lambda}$, equation (\ref{new00}) will be simplified to the equation (\ref{2Fried3}). Hence the two approaches are equivalent. It is also interesting to note that, in the approach explored by \cite{Emmanuel} by applying the first law of thermodynamics, the second Friedman equation is extracted, then by integration of this equation the first Friedman  is also obtained, however in approach explored by \cite{SheBFE}, by applying the first law of thermodynamics the first Friedman equation is extracted then by taking derivative of this equation and applying the conservation equation the second Friedman equation is also obtained \\
\section{Reconstructing the parameters of the model in terms of geometrical parameters}
In this section we aim to reconstruct the parameters of the model in terms of geometrical parameters. For simplicity, we define the following variables
\begin{equation}\label{new}
  \Omega_{m}=\frac{8\pi G_{\rm eff}\rho_{m}}{3H^{2}},\Omega_{\Lambda}=\frac{8\pi G_{\rm eff} \rho_{\Lambda}}{3H^{2}}
\end{equation}
Employing the above variables we can derive the following dynamical system\\

$\frac{d}{dx}
\left(\begin{array}{c}
  \Omega_{m} \\
  \Omega_{k} \\
  \Omega_{\Lambda}\\H \end{array}\right) =\left(\begin{array}{cccc}
                                   -1+2q & 0 & 0&0 \\
                                   0 &  2q& 0&0 \\
                                   0 & 0 & 2+2q&0\\
                                   0 & 0 & 0&-1-q
                                 \end{array}\right)
  \left(\begin{array}{c}
  \Omega_{m} \\
  \Omega_{k} \\
  \Omega_{\Lambda}\\H \end{array}\right)$
  \begin {equation}\label{dif}
.
\end {equation}
where $q$ is deceleration parameter and  $x=\ln a$. Also we  can use (\ref{new})
to rewrite the the first Friedman equation (\ref{Fried5}) as
\begin{equation}\label{om}
  \Omega_{m}+\Omega_{\Lambda}={H}^{-\Delta} \left( 1+\Omega_{k} \right) ^{1-\frac{1}{2}\,\Delta}
\end{equation}
Hence, by differentiating equation (\ref{om}) with respect to $x$ and using (\ref{dif}) , We can drive
\begin{align}
 &\Delta(1+q){H}^{-\Delta}+(1-2q){H}^{-\Delta} \left( 1+\Omega_{k} \right) ^{1-\frac{1}{2}\,\Delta} \nonumber\\
&+q(2-\Delta){H}^{-\Delta}\Omega_{k}\left( 1+\Omega_{k} \right) ^{-\frac{1}{2}\,\Delta}=3\Omega_{\Lambda}\label{o2}
\end{align}
In 1970, Alan Sandage \cite{Sandage} interpreted cosmology as the search for
two numbers: $ H_{0}$ and $q_{0}$
,then Weinberg \cite{Winberg} drew attention to the issue of extracting the value of constant spatial curvature $k$ and deceleration parameter $q$ from the observations,  without considering cosmological constant and/or scalar field. In 1976 Harisson \cite{Harrison} challenged Sandage's remark  and proved that the third derivative of the scale factor $Q$, is of great importance for observational cosmology, in a universe with indeterminate (dust) matter density.
He considered a universe containing the cosmological constant $\Lambda$
and non-relativistic matter. In this case, the Einstein equations reduce to the following
Friedmann equations
 \begin{align}\label{h0}
&H^{2}+\frac{kc^{2}}{a^{2}}=\frac{8\pi G}{3}\rho+\frac{\Lambda}{3}\\
&\dot{H}+H^{2}=-\frac{4\pi G}{3}(\rho+3P)+\frac{\Lambda}{3}\label{h00}
\end{align}
For zero-pressure model, the equations (\ref{h0}) and (\ref{h00}) can be combined as
 \begin{align}\label{h1}
K=4 \pi G\rho-H^{2}(q+1)
\end{align}
Where $K=\frac{kc^{2}}{a^{2}}$, $\rho$ is the average mass density.
The verification of the cosmological equation (\ref{h1}) requires the measurement of the quantities $K,H,\rho$ and $q$ \cite{Harrison}. In principle, these quantities $K,H,q$ can be determined, although in practice their precise determination is difficult \cite{Sandage}-\cite{Harrison}. If every thing could be known about the matter filling of the universe, then the cosmological constant will be derived as
  \begin{align}\label{h23}
\Lambda=4\pi G\rho- 3qH^{2}
\end{align}
Hence, for a universe with known amount of matter, the general relativity (GR) can be tested by measuring $H$ and $q$.  However, the  average density $\rho$ can not be determined, even in principal \cite{Sandage}-\cite{Harrison}. Here, the third derivative of the scale factor
is required to test the validity of the equation (\ref{h1}).
Since the parameters $(k, \Lambda,\rho)$  can be obtained in terms of first three derivatives of the scale factor as follows.
\begin{align}\label{h0000}
K=H^{2}(Q-1)\\
\Lambda=H^{2}(Q-2q)\\
4\pi G\rho=H^{2}(Q+q)\label{h0001}
\end{align}

It is easy to obtain the following relation
\begin{align}\label{qx}
\frac{dq}{dx}=-Q+q+2q^{2}
\end{align}
Hence, by differentiating equation (\ref{o2}) with respect to $x$ and using (\ref{dif}) and (\ref{qx}) , we can drive

\begin{align}\label{QQ}
(Q-1)= &\Omega_k-\Delta\,{q}^{2}\Omega_k\, \left( 1+\Omega_k \right) ^{-1}\nonumber\\&+  \left( 1
+2\,q+{q}^{2}+\Omega_k \right) \Delta
\end{align}
Hence, the Barrow parameter $\Delta$ can be obtained as
\begin{align}\label{dq}
\Delta=\frac{(Q-1-\Omega_k)(1+\Omega_k)}{(1+\Omega_k+q)^{2}}
 \end{align}
It is interesting to note that the relation (\ref{dq}) is established throughout the history of the Universe and is not limited to the present time, but we can use the information related to the current values of these parameters to determine the constant value of $\Delta$.
Based on the Planck 2018 results \cite{Planck 2018}, where, $\Omega_k=0.001\pm0.002$, $q_{0} =-0.527\pm0.01$ and the results of \cite{Nikodem} which have put constraint on parameter $Q_{0}$ as $Q_{0}=1.01^{+0.08}_{-0.021}$, the deviation relation shows that $\Delta =0.04^{+0.385}_{-0.068}$.
 In the the limiting case where $\Delta=0$, the equation (\ref{QQ}) reduces to
\begin{align}\label{oQQ}
\Omega_k=(Q-1)
\end{align}
Consequently using equation (\ref{oQQ}), for $\Delta=0$, the equations (\ref{om}) and (\ref{o2}) gives
\begin{align}\label{oQQ2}
\Omega_{\Lambda}=\frac{1}{3}(Q-2q)\\
\Omega_{m}=\frac{2}{3}(Q+q)\label{oQQ2}
\end{align}
Equations (\ref{oQQ})-(\ref{oQQ2}) are equivalent to equations (\ref{h0000})-(\ref{h0001}) which previously obtained by
\cite{Harrison}.
\subsection{Solution for case $\Omega_{k}=0$}
Considering $\Omega_k=0$, equation (\ref{dq}) gives the Barrow parameter $\Delta$ as
  \begin{align}\label{oQQb}
\Delta=\frac{Q-1}{(q+1)^{2}}
\end{align}

The first interesting result of equation (\ref{oQQb}) is that it indicates that there is profound connection between quantum-gravitational deformation effects $\Delta$ and geometrical cosmology parameters $(q,Q)$. Another interesting result is that this equation indicates that the geometrical parameters $(q,Q)$ can be regarded as deviation curvature factors which reflect the amount of deviation of the model from the standard model. The equation (\ref{oQQb}) indicates that for $Q=1$ the standard model $\Delta=0$ is recovered. This is an expected result because $Q=1$ is correspond to $\Lambda CDM$ model. In fact
almost all current cosmological observations can be summarized by a simple
case $(Q=1)$ \cite{Maciej},\cite{Alam},\cite{Kun}. Also Visser\cite{Visse}, have investigated in some details the condition $(Q=1)$.
In principal the case $Q=1$ is a third order ODE which is hold only for $\Lambda CDM$ model (It is easy to find that for flat case of  $\Lambda CDM$ model, $Q= 1+2\Omega_{r}$ , hence neglecting radiation density energy $\Omega_{r}=0$, $Q=1$). This relation can also be immediately deduced from Eq (\ref{h0000}).
In principle the parameter $Q$ as one of the pairs statefinder diagnostic $\{r\equiv Q,s\}$ can distinguish $\Lambda CDM$ form other cosmological models(For more discussion one can see (\cite{Sahni2003}  and \cite{Alam}).
The relation $Q=1$ exactly gives the evolution of the universe for $\Lambda CDM$ model. It gives the expected thermal history of the universe from matter dominated to dark energy dominated. To show this, we start from equation (\ref{qx}).By integration, it can be rewritten as
\begin{align}
\int dx=\int\frac{dq}{-Q+q+2q^{2}}
\end{align}
 Inserting $Q=1$, the above integral gives
$
x=\frac{1}{3}\ln(\frac{2q-1}{1+q})+\frac{1}{3}\ln(C)
$.
Here $ C$ is constant of integration.
 Since, $x=\ln a=-\ln(1+z)$, the above equation can be simplified as
 \begin{align}\label{qe404}
\ln\frac{1}{(1+z)^{3}}= \ln \Big(C(\frac{2q-1}{1+q})\Big)
 \end{align}
Hence, the deceleration parameter $q$ will be obtained in terms of redshift $z$ as
 \begin{align}\label{qe44}
q=\frac{C(1+z)^{3}+1}{2C(1+z)^{3}-1}
 \end{align}
By set $q_{0}$ for $z=0$ in this equation, we can obtain the constant $C$ in terms of $q_{0}$ as
$
C=\frac{1+q_{0}}{2q_{0}-1}
$. Also transition redshift $z_{t}$ will be obtained as
 \begin{align}\label{zt}
z_{t}=-1+(\frac{-1}{C})^{\frac{1}{3}}=-1+(\frac{1-2q_{0}}{1+q_{0}})^{\frac{1}{3}}
 \end{align}
According to the equation (\ref{qe44}), when $z\rightarrow\infty$, the deceleration parameter $q\rightarrow\frac{1}{2}$ and when $z\rightarrow-1$,
 the deceleration parameter $q\rightarrow-1$. Also if we get $q_{0}\simeq-0.53$ which is the result of Planck 2018, the equation (\ref{zt})  gives $z_{t}\simeq0.65$ \\
In Fig. 1, we have also plotted the evolution of deceleration parameter against redshift $z$ according to the equation (\ref{qe44}). As can be seen the evolution is exactly the evolution of the deceleration parameter for $\Lambda CDM$ model. Hence $Q=1$ is specified to flat(k=0) $\Lambda CDM$ by considering $\Omega_{r}=0$.
 \begin{figure}[t]
\includegraphics[scale=.4]{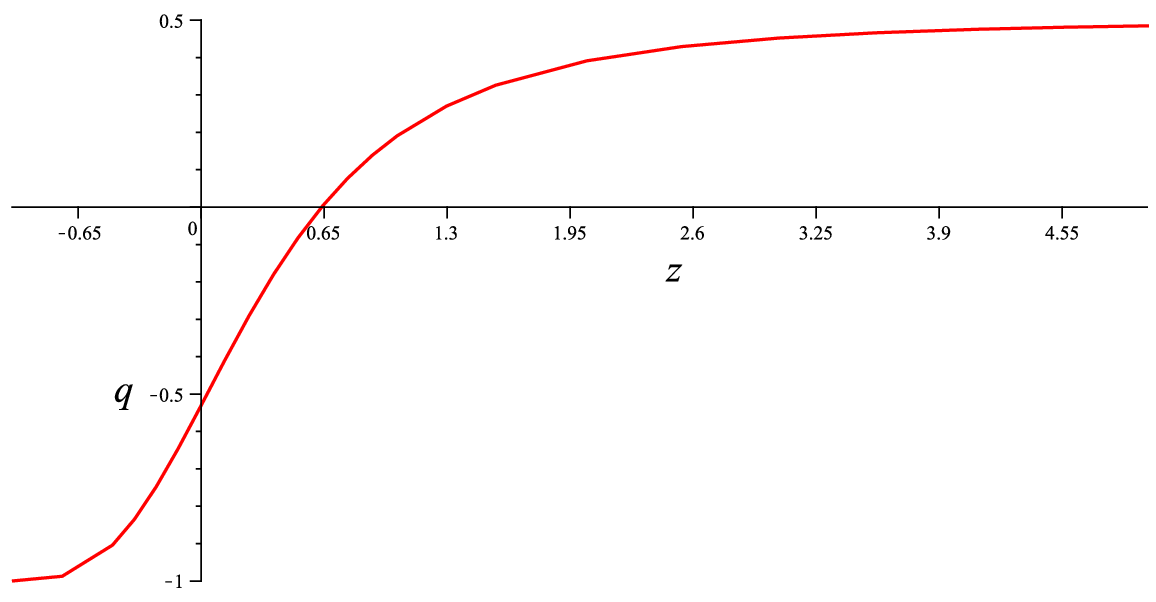}\hspace{0.1 cm}\\
Fig. 1:  The evolution of deceleration parameter against redshift $z$ according to equation (\ref{qe44})  \\
 \end{figure}\\
Therefore, it is expected that models whose structure is similar to the $\Lambda CDM$ model will have their Q closer to 1. This usually can be achieved in the models such as  Chevallier–Polarski–Linder (CPL), $\omega CDM$, $XCDM$ Phenomenologically Emergent Dark Energy (PEDE).
There are some previous studies which have also reported $Q\simeq1$  (i.e. the parameter $Q$ should be near  $1$).
 Moreover, there are some observational studies which have supported this expectation, $Q=1$. The Refs \cite{Maciej},\cite{Alam} find $Q=1$ and $Q=1.02$, respectively. Also, There are also some previous studies which have obtained values close to 1. For example see \cite{Herrera-Zamorano}, \cite{Escamilla}, \cite{Capozz}, \cite{Munoz} , \cite{Escamilla2}.
\cite{Harrison}.\\
In order to examine the sensitivity of the deviation parameter $\Delta$ with changes of each of these parameters, we introduce the following partial differential equations
\begin{align}\label{dev}
d_{Q}=\frac{\partial \Delta}{\partial Q},\ \
d_{q}=\frac{\partial \Delta}{\partial q}
\end{align}
Hence, using equations (\ref{oQQb}) and(\ref{dev}), we can find
\begin{align}\label{devi}
d_{Q}=&=\frac{1}{(q+1)^{2}}\\
d_{q}=&-2\frac{Q-1}{(q+1)^{3}}=-2\frac{\Delta}{q+1}
\end{align}
Hence, the absolute value of the ratio of these changes would be
\begin{align}\label{devi2}
\Big|\frac{d_{q}}{d_{Q}}\Big|=\Big|\frac{\Delta(q+1)}{2}\Big|
\end{align}
Since, $|q+1|<2$, one can deduced that
\begin{align}\label{ddq}
\Big|\frac{d_{q}}{d_{Q}}\Big|<\Delta
\end{align}
Due to this fact that $\Delta\ll1$, it can be concluded that $d_{q}\ll d_{Q}$. In other word, $\Delta$ is more sensitive to changes of $Q$.
 Hence, parameter $Q$ can be regarded as deviation parameter which determines the deviation of the model from $\Lambda CDM$. The more the difference of the $Q$ value is greater than 1, the greater the deviation of the model from the Standard model. Since, $0\leq\Delta\leq1$, so applying this condition on equation (\ref{oQQb}), a bound is obtained for parameter $Q$ as
  \begin{align}\label{oQQb2}
1\leq Q \leq 2+2q+q^{2}
\end{align}
Since, for accelerating Universe, $q_{0}<0$,
the condition (\ref{oQQb2})gives interesting result. It indicates that upper bound for $Q_{0}$ in Barrow cosmology is $Q_{0}<2$. This bound is further restricted  if we substitute the observational value of $q_{0}$. For $q_{0}\simeq-0.5$, which is the result of example for \cite{Planck 2018}, the condition is restricted to $1<Q_{0}<1.25$. This condition has interesting result as indicates that observational studies on Kaniadakkis cosmology must give the value for $Q_{0}$ which be closer to 1. It also indicates that as $q_{0}$ gets closer to 1, the region becomes more limited, where for $q_{0}=-1$ this region becomes a points as $Q_{0}=1$. This interesting result indicates that $\Delta=0$ is correspond to $(q_{0}=-1,Q_{0}=1)$ and these values are hold for $\Lambda CDM$ model. \\
Here we aim to find parameters of the model in terms of geometrical parameters $(H,q,Q)$.  Equation (\ref{o2}) for $\Omega_{k}=0$ can be rewritten as
\begin{eqnarray}\label{del20}
\Big(\Delta(1+q)+1-2q\Big)H^{-\Delta}=3\Omega_{\Lambda}
\end{eqnarray}
Where, inserting the $\Delta$ from equation (\ref{oQQb}) into the equation(\ref{del20}), the parameter, $\Omega_{\Lambda}$ will be obtained in terms of $(H,q,Q)$ as
\begin{eqnarray}\label{del21}
\Omega_{\Lambda}=\frac{1}{3}\Big(\frac{Q-q-2q^{2}}{q+1}\Big)H^{-(\frac{Q-1}{(q+1)^{2}})}
\end{eqnarray}
Also, inserting $H^{-\Delta}$, from equation (\ref{om}) into the equation(\ref{del20})
\begin{eqnarray}\label{dmm}
\Big(\Delta(1+q)+1-2q\Big)=\frac{3\Omega_{\Lambda}}{\Omega_{m}+\Omega_{\Lambda}}
\end{eqnarray}
Where, inserting the $\Delta$ from equation (\ref{oQQb}) into the equation(\ref{dmm}), we arrive
\begin{eqnarray}\label{ml}
\frac{\Omega_{m}}{\Omega_{\Lambda}}=-\frac{(Q-4q-2q^2-3)}{(Q-q-2q^2)}
\end{eqnarray}
Combining equations (\ref{del21}) and (\ref{ml}), the parameter $\Omega_{m}$ also can be obtained in terms of $(H,q,Q)$ as
\begin{eqnarray}\label{ml0}
\Omega_{m}=-\frac{1}{3}\Big(\frac{Q-4q-2q^2-3}{q+1}\Big)H^{-(\frac{Q-1}{(q+1)^{2}})}
\end{eqnarray}
The equations (\ref{oQQb}), (\ref{del21}) and (\ref{ml0}) are equations that express the parameters of the model,$(\Omega_{m},\Omega_{\Lambda},\Delta)$ in terms of geometrical parameters $(H,q,Q)$.\\
According to the approach explored by \cite{Emmanuel}, we can also find the interesting parameters $\Omega_{DE}$ and $\omega_{DE}$.
 For this peropus, we can
can re-express equations
(\ref{FRWgfe1}) and (\ref{firs}) as
\begin{eqnarray}
\label{FRWFR1}
&&H^2=\frac{8\pi G}{3}\left(\rho_m+\rho_{DE}\right)\\
&&\dot{H}=-4\pi G \left(\rho_m+p_m+\rho_{DE}+p_{DE}\right),
\label{FRWFR2}
\end{eqnarray}
 where
\begin{eqnarray}
&&
\!\!\!\!\!\!\!\!\!\!\!\!\!\!\!\!\!\!
\rho_{DE}=\frac{3}{8\pi G}
\left\{ \frac{\Lambda}{3}+H^2\left[1-\frac{ \beta (\Delta+2)}{2-\Delta}
H^{-\Delta}
\right]
\right\},
\label{FRWrhoDE1}
\end{eqnarray}
\begin{align}
&p_{DE}= -\frac{1}{8\pi G}\left\{
\Lambda+2\dot{H}\left[1-\beta\left(1+\frac{\Delta}{2}\right) H^{-\Delta}
\right]\right\}\nonumber\\+&3\left\{H^2\left[1- \frac{\beta(2+\Delta)}{2-\Delta}H^{-\Delta}
\right]
\right\}
\label{FRWpDE1}
\end{align}
where
respectively are the energy density and pressure of the  effective dark energy
sector
and $\beta\equiv \frac{4(4\pi)^{\Delta/2}G}{A_{0}^{1+\Delta/2}}$. Hence, by defining $\Omega_{DE}=\frac{8\pi G\rho_{DE}}{3H^{2}}$ and using equation (\ref{FRWrhoDE1}), we have
\begin{eqnarray}\label{oo}
\Omega_{DE}=\frac{8\pi G\rho_{DE}}{3H^{2}}=\Omega_{\Lambda}+\left[1-\frac{ \beta (\Delta+2)}{2-\Delta}
H^{-\Delta}
\right]
\end{eqnarray}
Also the effective equation of state read
 \begin{widetext}\begin{eqnarray}
 \omega_{DE}=\frac{p_{DE}}{\rho_{DE}}=\frac{-1}{3}\frac{\left\{
3\Omega_{\Lambda}
-2(1+q)\left[1-\beta\left(1+\frac{\Delta}{2}\right) H^{-\Delta}
\right]
+\left[1- \frac{\beta(2+\Delta)}{2-\Delta}H^{-\Delta}
\right]
\right\}}{\Omega_{\Lambda}+\left[1-\frac{ \beta (\Delta+2)}{2-\Delta}
H^{-\Delta}
\right]}
\end{eqnarray}
\end{widetext}
We, obtained $\Delta$ in terms of $\{q,Q\}$, also $\Omega_{\Lambda}$ and $\Omega_{m}$ in terms of cosmographic parameters $\{H,q,Q\}$, according to the above equations,
$\omega_{DE}$ and $\Omega_{DE}$ can also be expressed in terms of cosmographic parameters $\{H,q,Q\}$.\\
It is also possible to obtain the current value of parameters $\Omega_{m},\Omega_{\Lambda},\Omega_{DE},\omega_{DE}$ in terms of the current values  $(H_{0},q_{0},Q_{0})$.
Here, we aim to test the model for different cosmographic sets. It is important to note that although, the present observational data could provide strong constraint on the current values of Hubble parameter $H_{0}$ and deceleration parameter $q_{0}$, but due to the fact that constraining on $Q_{0}$ requires high redshift data and until today, there is no enough high-quality data at high redshift,it is not possible to put strong constrain on $Q_{0}$ and different studies have reported different values for this parameter. Hence there is no very reliable set of $\{H_{0},q_{0},Q_{0}\}$. However, it is possible to estimate this parameter theoretically in different cosmological models.
We can use the results of Planck 2018 \cite{Planck 2018} for $H_{0}$ and $q_{0}$ as robust observational measurements. According to the Planck results, $H_{0}=(67.4\pm0.5)$ and $q_{0} =-0.527\pm0.01$. However, since no strong measurements on $Q_{0}$ and $\Delta$ and wide ranges of values have been reported for this parameters. For example authors of \cite{Alejandro}, using same model, reported different values  $Q_{0}\simeq1.268$ and $Q_{0}\simeq -7.746$ for different data. In the \cite{Alejandro}, they also reported $Q_{0}\simeq-13.695$ for another category of the model. So there is no strong  constrain on $Q_{0}$. We are also facing this problem for the parameter $\Delta$. Recently, using Big Bang Nucleosynthesis (BBN) data \cite{John} have imposed constraint on the exponent $\Delta$ and found that Barrow exponent should be inside the bound $\Delta\leq 1.4\times 10^{-4}$, also the authors of \cite{Genly}, found the value $\Delta=5.912\times10^{-4}$ which is comparable with\cite{John}. The authors of \cite{Mahnaz}, found $\Delta\simeq10^{-4}$.
In \cite{Leon} the entropic modified Friedmann equation within the gravity-thermodynamics approach is confronted with a set of cosmological probes, including Pantheon SNeIa and a BAO sample, and obtain $\Delta \sim 10^{-4}$.
In \cite{{Feizi}} the authors do not use any early-times probe, but only SNeIa, CC and GRBs, and a different horizon from ours, and their final estimation for the Tsallis parameter is $\delta \approx 0.16$, which corresponds to $\Delta \approx -1.68$, clearly out of the physical boundary required by Barrow theory. In \cite{S0} the holographic principle is applied and they found the Tesallis parameter as $\delta \approx 1.07$ corresponding to $\Delta \approx 0.14$. Similar results are obtained in \cite{S1,S2} using only late-times data, with $\Delta \sim 0.09$. Finally, in \cite{{Adhikary}}, the Barrow entropy parameter was constrained as  $\Delta \sim 0.03 $. The authors of\cite{Nandhida}, using Barrow Holographic Dark Energy Model with GO Cut-of found $\Delta = 0.063\pm 0.029$\\
We see that different values from order $10^{-4}$ to order $1$ have been reported for parameter $\Delta$. Hence same as $Q_{0}$, there is no convergence in results.\\
Here, according to the analytical relations which have obtained, we want to test the different results.\\
At first we call the relation (\ref{oQQb}). Since, for accelerating universe $q_{0}<0$, hence, $(q_{0}+1)<1$ and ($(q_{0}+1)^{2}<1$
 Hence according to the equation (\ref{oQQb})
 \begin{align}
 (Q_{0}-1)\leq\Delta
 \end{align}
 This condition indicates that for those studies that reported $\Delta\simeq10^{-4}$, the condition $(Q_{0}-1)<10^{-4}$ should be satisfied..
This indicates that $Q_{0}$ should be very close to $1$. It also points out an important point; Observational constraining on $Q_{0}$
should be performed with high accuracy and sensitivity, so that the error of the order of $10^{-4}$ is measured. Although, there are many studies that have reported best fitted values for $Q_{0}$ which are close to 1, but in none of them, a best fitted value less than $1.01$ has been reported. Among all the values reported for $Q_{0}$, the closest value to 1 has been reported by \cite{Nikodem}, which is $Q_{0}=1.01^{+0.08}_{-0.021}$.
Considering this value
 and result of Planck 2018\cite{Planck 2018} for $q_{0}$; $q_{0} =-0.527\pm0.01$, the relation
gives $\Delta =0.04^{+0.385}_{-0.068}$. Although this result is consistence with some of previous studies, however, the other parameters should be also measured to test the validation of the result.
Considering the Planck 2018 result\cite{Planck 2018} $H_{0}=(67.4\pm0.5)$, the current values of parameters of the model for $\Delta =0.04^{+0.385}_{-0.068}$, are reconstructed as
\begin{align}
&\Omega_{m0}=.255,\ \ \Omega_{\Lambda_{0}}=0.573, \nonumber\\ &\Omega_{DE_{0}}=0.07,\ \ \omega_{DE0}\simeq-8.36 \nonumber\\
&\beta\simeq1.73
\end{align}
 As can be seen, the values obtained for $\Omega_{DE0},\omega_{DE0}$ are not consistence with that expected by robust observational data and indicates that the value obtained for $\Delta$ is not desired value. This is due to this fact that parameter $\Delta$ should be very close to 0, and this requires the value of $Q_{0}$ is very close to 1.
If we check the model for upper bound of $\Delta$ which reported by \cite{John}; $\Delta= 1.4\times 10^{-4}$ and considering the Planck results $q_{0} =-0.527\pm0.01$,$H_{0}=(67.4\pm0.5)$, the results would be
\begin{align}
&\Omega_{m0}\simeq0.3151,\ \ \Omega_{\Lambda_{0}}\simeq0.6842, \nonumber\\ & \Omega_{DE_{0}}\simeq0.6830,\ \ \omega_{DE0}\simeq-1.0014,\nonumber\\
&\beta\simeq 1.00122
\end{align}
The values obtained for this case are in excellent agreement with observations.
  In particular the value $\Omega_{m0}\simeq0.3151$ is very close to that obtained by \cite{Planck 2018} as $\Omega_{m0}\simeq0.315\pm0.007$.
 According to the equation (\ref{oQQb}) for $\Delta= 1.4\times 10^{-4}$ and $q_{0} =-0.527\pm0.01$ the current value $Q_{0}$ is  \begin{align}(Q_{0}\simeq1.00002)\end{align}
 If we test the model for $\Delta=5.912\times10^{-4}$ which obtained by \cite{Genly}, the following results would be obtained
 \begin{align}
&\Omega_{m0}\simeq0.3144,\ \ \Omega_{\Lambda_{0}}\simeq0.6830, \nonumber\\ & \Omega_{DE_{0}}\simeq0.6776,\ \ \omega_{DE0}\simeq-1.0080,\nonumber\\
&\beta\simeq 1.00122
\end{align}
These results are very close to results of \cite{Genly}, who found
\begin{align}
&\Omega_{m0}=311^{+0.006}_{-0.005},\ \omega_{DE0}\simeq-1.0001,
\beta=0.920^{+0.042}_{-0.042}
\end{align}
For this case the current value of $Q_{0}$ be
\begin{align}(Q_{0}\simeq1.00012)\end{align}
 However, as we mentioned, so far, the measurement on $Q_{0}$ has not been done with this high accuracy.
  The fact that values closer to 1 are not reported may be due to this fact that, so far, the importance of very small changes
   of order $10^{-4}$ of this parameter and how much it can affect the results has not been paid attention to. So  observational measurements and numerical analysis on $Q_{0}$ are not done with this accuracy. As mentioned, another reason could be the lack of high quality data required for constraining this parameter.
 So this study indicates the parameter $Q$ is important parameter in Barrow campanology which its current value is very close to 1. In Table(\ref{tmodel}), we have obtained current values for parameters of the model, for different values of $Q_{0}$ by fixing Planck 2018 results for $q_{0}$ and $H_{0}$.
Due to the high sensitivity of the model with small changes of this parameter, even very small errors in estimation of this parameter lead to very different results for the model. As we investigated, very different results was obtained for two very close values $(Q_{0}\simeq1.01)$ and $(Q_{0}\simeq1.001).$
\begin{table}
\caption{\label{tmodel} Reconstructing parameters for  diffferent values of $Q_{0}$ by fixing $q_{0}=-0.527$ }
%\begin{center}
\begin{tabular}{ccccccc}
%\hline
$Q_{0}$  & $\Delta$& $\Omega_{m0}$  &$\Omega_{\Lambda0}$ \ & $\Omega_{DE0}$ \ & $\omega_{DE0}$\ & $\beta$  \\
% inserting body of the table
\hline % inserts single horizontal line
\hline
$1$  &0& 0.3155 &$ 0.6846 $& 0.6846&-1&1
\\
% inserting body of the table
\hline$1.00001$ &0.00004 &0.3152 &$ 0.6845 $& 0.6841&-1.0006&1.00054 \\
% inserting body of the table
\hline$1.0001$ &0.000044 &0.31467 &$.6834 $&.6793&-1.0066&1.0055
\\
% inserting body of the table
\hline$1.001$ &0.0044 &0.30876 &$ 0.67259 $& 0.63118&-1.0725&1.05645 \\
% inserting body of the table
\hline$1.005$ &0.0223 &0.2838 &$0.6263 $& 0.4014&-1.6098&1.31601 \\
% inserting body of the table
\hline$1.008$ &0.0357 &0.2664 &$0.5938 $&0.2103&-2.9622&1.5517 \\
% inserting body of the table
\hline$1.01$  &0.0446& 0.2553 &$ 0.5730 $&0.0726&-8.36038&1.7318
\\
% inserting body of the table
\hline$1.1$ & 0.446 &0.0372 &$0.11499 $&-57.13&-0.0831&242.77 \\
% inserting body of the table
\hline % inserts single horizontal line
\hline\end{tabular}
%\end{center}
\end{table}
\subsection{Solution for case $\Omega_{k}\neq0$}
 For general case $(\Delta\neq 0,\Omega_k\neq0)$, from equation(\ref{QQ}), $\Omega_k$ can be obtained as
\begin{equation}\label{omk}
\Omega_k=-1+{\frac {-2\,\Delta\,q+Q\pm\Big({Q}^{2}-4\,\Delta\,qQ-4\,\Delta\,
{q}^{2}\Big)^{\frac{1}{2}}}{2(1+\Delta)}},
 \end{equation}
 By substituting $\Omega_k$ from equation (\ref{omk}) in equation (\ref{o2}), $\Omega_\Lambda$ is obtained in terms of $(H,q,Q,\Delta)$.
In equation (\ref{dq}), we have obtained $\Delta$ in terms of the cosmographic parameters $(q,Q)$ and $\Omega_k$, however if we want to obtain $\Delta$ only in terms of the cosmographic parameters and eliminate the parameter $\Omega_k$, we need the cosmographic parameter $X=\frac{\ddddot{a}}{aH^{4}}$.  For this respect, by taking derivative of both sides of equation (\ref{QQ})
and using equation (\ref{dif})
we can obtain
\begin{align}\label{Qx2}
&-4\Delta q^3+(-8\Delta\Omega_k-6\Delta)q^2+\Big((-4\Delta-4)\Omega_k^2 \nonumber\\
&+(5Q-6\Delta-4)\Omega_k+(2Q-2)\Delta+3Q\Big)q\nonumber\\
&+(1+\Omega_k)(X+2Q+12+2\Delta Q)=0
\end{align}
Where we have also used the following relation
\begin{align}\label{Qx}
\frac{dQ}{dx}=X+(2+3q)Q
\end{align}
From equations (\ref{Qx2}) and (\ref{omk}), the barrow parameter $\Delta$ can be obtained in term of the cosmographic parameters $(q,Q,X)$ as

\begin{align}\label{X0}
&\Delta=-\frac{1}{4(q+Q)^{3}}\Big(2AQ+\frac{1}{2}AqQ+2Aq+6A+\frac{1}{2}A \nonumber\\+&(qQ^2+XQ+12Q)(q+Q)^{3}\Big)
\end{align}
Where
\begin{align}\label{XX}
A=&-4qQ-q^2Q-4q^2-12q-qX\nonumber\\
&+\Big(144q^2+12XQq^2+56q^3Q+2q^3QX \nonumber\\
&+q^4Q^2+8q^4Q+q^2X^2+144q^2Q+96q^3\nonumber\\
&+20q^2Q^2+12q^3Q^2+24Xq^2+8Xq^3+16q^4\nonumber\\
&+8qQ^3+4Q^4+4q^2Q^3+4qXQ^2+48qQ^2\Big)^{\frac{1}{2}}
\end{align}
We have used Maple software to derive the above relations.
By substituting $\Delta$ from equation (\ref{X0}) into equation (\ref{omk}), the parameter $\Omega_k$ is also obtained in terms of cosmographic parameters $(q,Q,X)$.
Also, from equation (\ref{o2}),  the parameter $\Omega_\Lambda$ is obtained in terms of $(H,q,Q,X)$, finally from equation (\ref{om}), the parameter $\Omega_m$ is also obtained in terms of $(H,q,Q,X)$. This means that it is possible to reconstruct the parameter of the Barrow model $(\Delta,\Omega_k,\Omega_m,\Omega_\Lambda)$ in terms of directly measurable parameters $(H,q,Q,X)$.

\section{summary and remarks}
In this paper we reconstructed the parameters of the modified Friedman equation due to the gravity-thermodynamics approach and assuming the Barrow entropy in terms of the first four cosmographic parameters $(H,q,Q,X)$. Barrow entropy arises from the fact that the black-hole surface may
be deformed due to quantum-gravitational effects, and its
deviation from Bekenstein-Hawking one is quantified by $\Delta$. Due to significant role of the parameter $\Delta$ in describing the late-time
universe and early-time behavior, most of the recent studies on Barrow entropy have focused on imposing constraints on
the exponent $\Delta$ . We found that for a universe filled with mater and cosmological constant $\Lambda$ this parameter could be obtained in terms of $\Omega_{k}$ and  two important cosmographic parameters $(q,Q)$ where for flat universe or neglecting $\Omega_{k}$ it can be obtained only in terms of $(q,Q)$. The connection between parameter $\Delta$ which reflects the quantum-gravitational effects and cosmographic parameters which reflect the  curvature effects has interesting features which can be summarized as follows.\\
The first interesting feature is that this relation indicates that it is possible to imposing constrain on parameter $\Delta$ by imposing constrain on parameters current values of $(q,Q)$. This can be interesting because of this facts that cosmographic parameters $(q_{0},Q_{0})$ could be measured directly without need of any background cosmological model, in addition one of the main purposes of observational cosmology is to get precise measurements of the first cosmographic parameters $(H_{0},q_{0},Q_{0})$ that will provide a crucial test for cosmological models \cite{Sandage},\cite{Harrison}. Hence a vast majority of cosmological observational studies have focused on constraining on these parameters. As a robust observational data, the  Planck 2018 \cite{Planck 2018} providing a major source of information relevant to many cosmological parameters
 indicates that $q_{0} =-0.527\pm0.01$. Also \cite{Harrison} proved that the third derivative of the scale factor $Q$, is of great importance for observational cosmology, hence some of studies only have focused on constraining cosmic jerk parameter $Q$ in different cosmological models. The authors of \cite{Nikodem} have put constrain on parameter $Q_{0}$ in $f(R)$ gravity and find $Q_{0}=1.01^{+0.08}_{-0.021}$. Considering this value and the value of $q_{0} =-0.527\pm0.01$ obtained from Planck 2018 results the deviation parameter $\Delta$ is obtained as $\Delta =0.04^{+0.385}_{-0.068}$.
 We showed that although this result is consistence with some of previous studies, however, the other important parameters should be also measured to test the validation of the result. We reconstructed important parameters $\Omega_{DE_{0}}, omega_{DE0}$ and $\beta$ according to the approach explored by \cite{Emmanuel} and test the model by this parameter.
For $\Delta =0.04^{+0.385}_{-0.068}$, considering the Planck 2018 result\cite{Planck 2018} $H_{0}=(67.4\pm0.5)$, the current values of parameters of the model for this value, are reconstructed as $\Omega_{DE_{0}}=0.07, \omega_{DE0}\simeq-8.36$
which are not consistence with that expected by robust observational data and indicates that the value obtained for $\Delta$ is not desired value. This is due to this fact that parameter $\Delta$ should be very close to 0, and this requires the value of $Q_{0}$ be very close to 1.
 We checked the model for upper bound of $\Delta$ which reported by \cite{John}; $\Delta= 1.4\times 10^{-4}$ and considering the Planck results $q_{0} =-0.527\pm0.01$,$H_{0}=(67.4\pm0.5)$, and
\begin{align}
\Omega_{m0}\simeq0.315,\ \Omega_{\Lambda_{0}}\simeq0.684,  \Omega_{DE_{0}}\simeq0.683, \omega_{DE0}\simeq-1.0014,\nonumber
\end{align}
Also $\beta$ was found as $\beta\simeq 1.00122$. The values obtained for this case are in excellent agreement with observations.
  In particular the value $\Omega_{m0}\simeq0.3151$ is very close to that obtained by \cite{Planck 2018} as $\Omega_{m0}\simeq0.315\pm0.007$.
 According to the equation (\ref{oQQb}) for $\Delta= 1.4\times 10^{-4}$ and $q_{0} =-0.527\pm0.01$ the current value $Q_{0}$ is $(Q_{0}\simeq1.00002).$
 We also test the model for $\Delta=5.912\times10^{-4}$ which obtained by \cite{Genly}, the following results would be obtained
 \begin{align}
&\Omega_{m0}\simeq0.314,\ \ \Omega_{\Lambda_{0}}\simeq0.683, & \Omega_{DE_{0}}\simeq0.677,\ \ \omega_{DE0}\simeq-1.008 \nonumber
\end{align}
For this case the parameter $\beta$ was found as $\beta\simeq 1.00122$. These results are very close to results of \cite{Genly}, who found
\begin{align}
&\Omega_{m0}=311^{+0.006}_{-0.005},\ \omega_{DE0}\simeq-1.0001,
\beta=0.920^{+0.042}_{-0.042}\nonumber
\end{align}
For this case the current value of $Q_{0}$ be
$(Q_{0}\simeq1.00012)$.
 However, so far, the measurement on $Q_{0}$ has not been done with this high accuracy.
  The fact that values closer to 1 are not reported may be due to this fact that, so far, the importance of very small changes
   of order $10^{-4}$ of this parameter and how much it can affect the results has not been paid attention to. So  observational measurements and numerical analysis on $Q_{0}$ are not done with this accuracy. Another reason could be the lack of hat high redshift there is no strong constrain on this parameter.\\
 So, we can conclude that this study reveals the importance of parameter $Q$ in Barrow campanology, it also indicates that the current value of $Q_{0}$ is very close to 1.
   Our theoretical approach indicates that,
 the works done by \cite{John} and \cite{Genly} which found that Barrow exponent should be inside the bound $\Delta\leq 1.4\times 10^{-4}$ have good agrement with observations. In other word for this bound, the values which reconstructed for parameters of the model are in excellent agreement wit observations. This bound indicates that by considering Planck results \cite{Planck 2018}, for $q_{0} =-0.527\pm0.01$ and $\Omega_k=0.001\pm0.002$ the relation (\ref{dq}) predicts the value for $Q_{0}$ which is slightly deviates from 1 as $(Q_{0}-1)<0.001$. This can be a relativity well target and criterion for theoretical and observational measurements of parameter $Q$, so we can hope and wait the improvement of the high redshift data in the future to support it.

\section{Data availability statement}

The manuscript has no associated data or the data will not be deposited

\end{document}